\documentclass{ws-ijmpe}
\usepackage[super]{cite}
\usepackage{xcolor}
\usepackage[verbose,hypertexnames=false]{hyperref}
\hypersetup{colorlinks=false,allbordercolors=blue,pdfborderstyle={/S/U/W 1}}

\begin{document}

\markboth{Authors' Names}{Instructions for typing manuscripts (paper's title)}

\catchline{}{}{}{}{}

\title{The nuclear surface diffuseness effects on the alpha decay of heavy and super heavy nuclei}

\author{S. Mohammadi$^{1}$, R. Gharaei$^{2}$\footnote{Email: rezagharaei@um.ac.ir},  S. A. Alavi$^{1}$}

\address{$^{1}$Department of Physics, Sciences Faculty, Hakim Sabzevari University, P. O. Box 397, Sabzevar, Iran
	\\$^{2}$Department of Physics, Faculty of Science, Ferdowsi University of Mashhad, P.O. Box 91775-1436, Mashhad, Iran}
\maketitle

\begin{history}
\received{Day Month Year}
\revised{Day Month Year}
\accepted{Day Month Year}
\published{Day Month Year}
\end{history}

\begin{abstract}
The present study aims at investigating systematically the effects of nuclear surface diffuseness on the alpha decay within the framework of the Coulomb and proximity potential model along with the WKB approximation. We select 300 different parent nuclei in the range $Z=64-106$. The proximity potentials Zhang 2013 and Guo 2013 are employed to calculate the nuclear potential. The influence of the nuclear surface diffuseness is applied in the calculations of interaction potential between the emitted alpha particle and daughter nucleus through the reduced radius parameter $\bar{C}$. The systematic analysis of the radial behavior of interaction potential with and without the surface diffuseness effect reveals that these effects play decisive role in the calculation of nucleus-nucleus potential at the touching radius of the two interacting nuclei. We indicate that its influence decreases outside the touching configuration. In addition, our results reveal that the barrier penetration probability of the alpha particle through the barrier decreases by imposing the mentioned physical effects. It is worth noting that the calculatd $\alpha$-decay half-lives using the Zhang 2013 and Guo 2013 proximity potentials accompanied by the surface effects agree very well with the available experimental data. The theoretical half-lives are calculated for 50 super-heavy nuclei using the modified forms of the Zhang 2013 and Guo 2013 models. The comparison with available experimental data and also with different empirical formulas demonstrates that the Guo 2013 model is suitable to deal with the alpha decay half-lives of SHN. Then the predictions of alpha decay half-lives for 65 SHN with $Z = 120-126$ are made by using Guo 2013 model with the surface effects. We found that there is a good agreement between our predicted half-lives and those obtained from semi-empirical formulas such as Royer and UDL.
\end{abstract}


\ccode{PACS Number(s): 23.60.+e, 21.10.Tg, 24.10.-i\\}

\section{Inroduction}\label{1}
In nuclear physics studies, alpha-radioactivity is known as one of the most important decay modes for unstable medium, heavy and super-heavy nuclei. In the early 20th century, this process was theoretically introduced as a quantum tunneling process. In order to solve analytically the problem of the penetration of an alpha particle through the barrier, the Wentzel-Kramers-Brillouin (WKB) approximation was found to be a suitable approximation. In the framework of this semi-classical formulation, the total interaction potential between $\alpha$ particle and daughter nucleus plays an important role. It is well-known that this potential consists of three components: the short-range attractive nuclear potential, the long-range repulsive electrostatic potential and the centrifugal potential. The Coulomb and centrifugal terms are well recognized, whereas there are large ambiguities in the optimum form of the nuclear potential. So it is necessary to choose an appropriate model to determine this part of total interaction potential. During recent decades, the different phenomenological, microscopic and macroscopic potential models have been introduced \cite{1,2,3}. The phenomenological proximity potential is one of the most widely used theoretical approaches in alpha-decay studies \cite{1}. This model is based on the proximity force theorem. Due to different adjustable parameters, the proximity model can expand and develop. Therefore, various modified versions of the proximity potentials have been presented so far\cite{4,5,6,7,8,9,10,11,12,13,14,15,16,17}. In 2013, Guo {\it et al.}\cite{16} and Zhang {\it et al.}\cite{17} introduced a new form of the universal function of proximity potential using the double-folding model with the density-dependent nucleon-nucleon interaction of CDM3Y6-type\cite{2}. We mark these proximity potentials as "Zhang 2013" and "Guo 2013", respectively. 

During recent years, the authors analyzed the influence of various physical effects such as the coupled-channel effects, the temperature effects of parent nuclei and also the nuclear surface tension effects on the alpha-decay process\cite{18,19,20,21,22,23}. For example, Zanganeh and coworkers investigated the role of temperature dependence of the interaction potential in the half-lives of alpha and cluster decays for Fr isotopes\cite{24}. They indicated that the probability of decay increases with the increase of nuclear temperature. Nuclear surface diffuseness (NSD) is another important physical property which is related to Fermi-level nucleon occupancy\cite{25}. In fact, this effect appears as a result of the specific distribution of nuclear matter and the short-range nuclear force of valence nucleons. In 1992, Gupta and coworkers investigated the role of NSD in the estimated half-life for exotic cluster decays by using the preformed cluster model.\cite{26} The authors found that the effects of the diffuseness of the nuclear surface are important for the proper interpretation of the cluster preformation probabilities and the barrier penetration probabilities. They also indicated that these effects become smaller for lifetime estimation as the mass number of the emitted cluster becomes larger than about 20. In 2010, Dutt and Puri discussed the impact of nuclear surface diffuseness in the potential and ultimately in the fusion process of heavy-ions.\cite{11} The obtained results revealed that NSD can affect the nuclear potential as well as fusion barriers. In the present work, we are interested in studying the influence of the nuclear surface diffusion on the different characteristics of alpha-decay process, such as the potential barrier between $\alpha$-particle and daughter nucleus and alpha-decay half-lives. 

This paper is organized as follows. The details of the calculations of the total interaction potential are presented in Sec.~\ref{2}. In Sec.~\ref{3}, we present our parameterization method. A comparison with the results of the experimental data made in this sections. The Sec.~\ref{4} is dedicated to the summary and concluding remarks.

\section{Theoretical Framework}\label{2}
The alpha radioactivity half-life is related to the decay constant $\lambda$ as,
\begin{equation} \label{T12}
	T_{1/2}=\frac{\rm ln 2}{\lambda},
\end{equation}
where the decay constant $\lambda$ can be calculated using the following definition,
\begin{equation} \label{lam}
	\lambda=\nu P_0 P,
\end{equation}
here $\nu$ is known as the assault frequency (refers to the number of alpha particle attacks on the barrier per second).
\begin{equation} \label{DFinte}
	\nu = \frac{\omega}{2\pi}=\frac{2E_{\nu}}{h},
\end{equation}
where $h$ and $E_{\nu}$ are Planck constant and the empirical vibrational energy, respectively.\cite{27} In Eq.~(\ref{lam}), the $\alpha$ preformation factor $P_0$ is an indispensable quantity for the calculation and can supply the information of the nuclear structure. It is obtained as 0.43 for even-even nuclei, 0.35 for odd-A nuclei and 0.18 for odd-odd nuclei.\cite{28} Within the framework of the WKB approximation, one can calculate penetration probability $P$ as follows
\begin{equation} \label{pene}
	P=\textmd{exp}\bigg[-\frac{2}{\hbar}\int_{R_a}^{R_b}\sqrt{2\mu
		(V_{\rm tot}(r)-Q_{\alpha})}dr\bigg],
\end{equation}
where $V_{\rm tot}$, $Q_{\alpha}$, and $\mu=\frac{m_{\alpha} m_d}{m_{\alpha}+m_d}$ are the total interaction potential, the released energy of the emitted $\alpha$-particle, and the reduced mass of the two-body system, respectively. Moreover, in Eq.~(\ref{pene}), $R_a$ and $R_b$ refer to the physical turning points and are given by,
\begin{equation} \label{cond}
	V_{\rm tot}(R_a)=Q_{\alpha}=V_{\rm tot}(R_b).
\end{equation}
We know that the total interaction potential in the alpha-nucleus system consists of three parts: $V_C$ (Coulomb), $V_{\ell}$ (centrifugal) and $V_N$ (nuclear) potentials.
\begin{equation} \label{tot}
	V_{\rm tot} (r)= V_C (r)+V_{\ell} (r)+V_N (r),
\end{equation}
here $r$ is the separation distance between the centers of mass of the alpha-daughter system. To calculate the Coulomb potential, the following quasi-point approach can be used, in which the calculations are performed with the assumption that the nuclei participating in the interaction are uniformly charged spheres,
\begin{eqnarray}\label{col}
	V_C(r) = Z_{\alpha}Z_de^2\left\{\begin{array}{rl}
		\frac{1}{r} ~~~~~~~~~~~~~~~~~~~~~~~~~r\geq R_C \\
		\\
		\frac{1}{2R_C}\Bigg[3-\bigg(\frac{r}{R_C}\bigg)^2\Bigg]~~~~ r\leq R_C\\
	\end{array} \right.
\end{eqnarray}
where $Z_i$ refers to the alpha particle and daughter nucleus charge number. Moreover, the $R_C$ parameter is equal to the sum of the interacting nuclei radii (namely $R_C=R_{\alpha}+R_d$). Centrifugal potential with angular momentum $\ell$ carried by alpha particle is calculated as follows,
\begin{equation} \label{vl}
	V_{\ell} (r)=\frac{\ell(\ell+1)\hbar^2}{2\mu r^2}.
\end{equation}
In order to calculate the nuclear part of the interaction potential, Guo 2013\cite{16} and Zhang 2013\cite{17} potential models are introduced. These models consist of two basic parts: the first part includes factors that depend on the shape and geometry of the interacting nuclei. The second part contains a universal function that depends only on the separation distance between the surfaces of two interacting nuclei. The nuclear potential is calculated as, 
\begin{equation} \label{VN}
	V_{N}(r)=4\pi b \gamma \overline{R}\Phi\bigg(\frac{s}{b}\bigg),
\end{equation}
where $b$ is the surface thickness parameter with an approximate value of 1 fm. $\bar{R}$ is the average curvature radius and can be defined as,
\begin{equation} \label{Rbar}
	\overline{R}=\frac{R_{\alpha}R_d}{R_{\alpha}+R_d},
\end{equation}
here $R_i$ refers to the effective sharp radius of the $\alpha$-daughter system which is given as,
\begin{equation} \label{Ri}
	R_i=1.28A_i^{1/3}-0.76+0.8A_i^{-1/3},~~   (i=\alpha, d).
\end{equation}
In Eq.~(\ref{VN}), $\gamma$ is the surface energy coefficient and has the form
\begin{equation} \label{gamma}
	\gamma=\gamma_0\Bigg[1-k_s\bigg(\frac{N-Z}{N+Z}\bigg)^2\Bigg],
\end{equation}
herer $N$ and $Z$ represent the neutron and proton numbers of parent nucles, respectively. Moreover, $\gamma_0$ and $k_s$ coefficients are the surface energy and surface asymmetry constants. In the original proximity version \cite{1}, the values of these constants are equal to 0.9517 MeV.$\rm fm^{-2}$ and 1.7826, respectively. The universal function $\Phi(s=r-R_1-R_2)$ used in Eq.~(\ref{VN}) depends only on the surface separation distance of two interacting nuclei,
\begin{equation} \label{phi}
	\Phi(s)=\frac{p_1}{1+{\rm exp}(\frac{s+p_2}{p_3})},
\end{equation}
where the constant values of $(p_1, p_2, p_3)$ for the Guo 2013 and Zhang 2013 proximity potentials are reported as (-17.72, 1.30, 0.854)\cite{16} and (-7.65, 1.02, 0.89)\cite{17}, respectively.
 
\section{Results and Discussion}\label{3}
\subsection{Alpha decay half-lives in heavy nuclei region}
\begin{figure}[th]
	\centerline{\includegraphics[width=11cm]{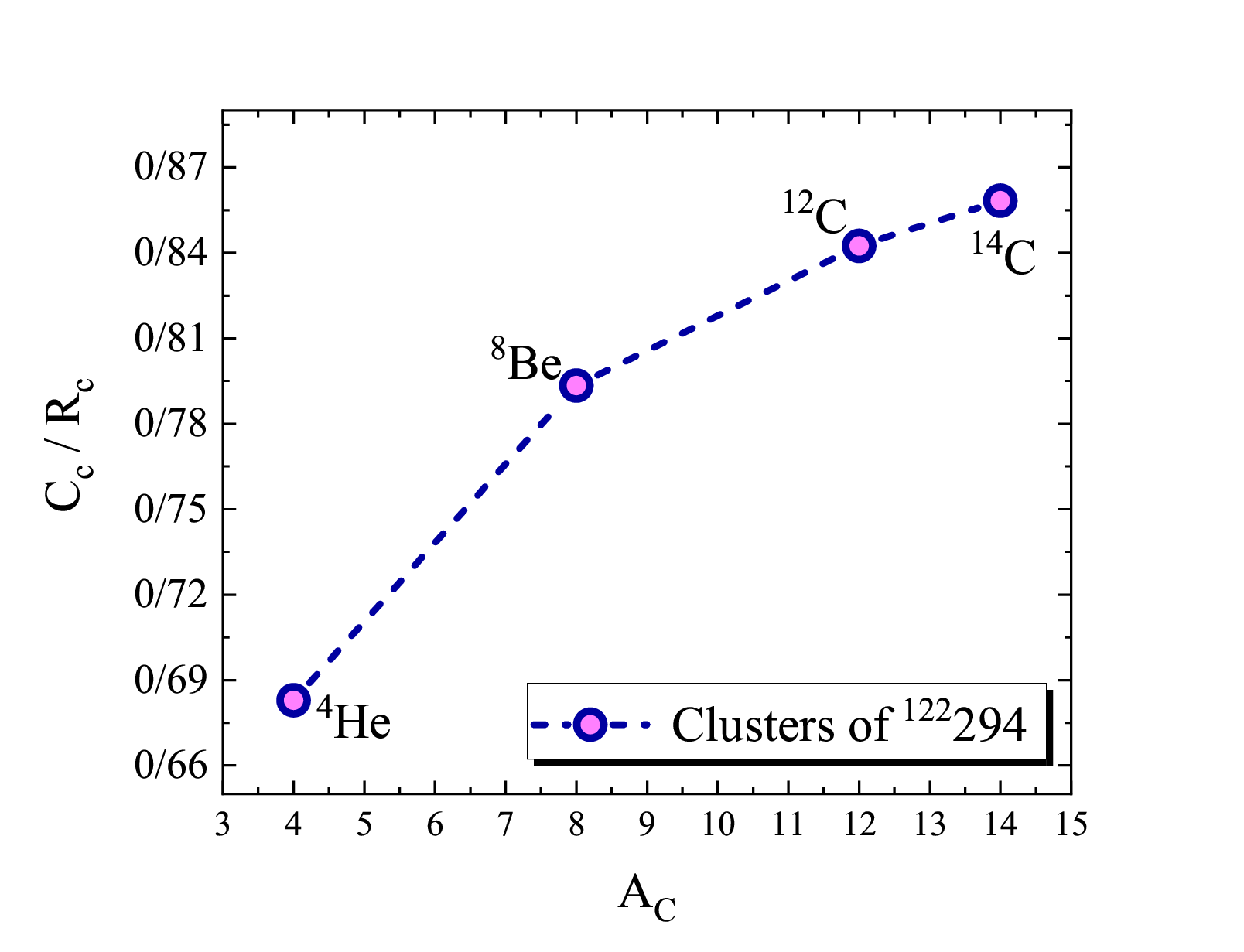}}
	\caption{\footnotesize (Colored online) The ratio $\frac{\bar{C}}{\bar{R}}$ (the central and sharp radii) as a function of the mass number of the $^4$He, $^8$Be, $^{12}$C, and $^{14}$C clusters of the $^{122}$294.}
	\label{CbarRbar}
	\alttext{Description of the figure}
\end{figure}

\begin{figure}[th]
	\centerline{\includegraphics[width=11cm]{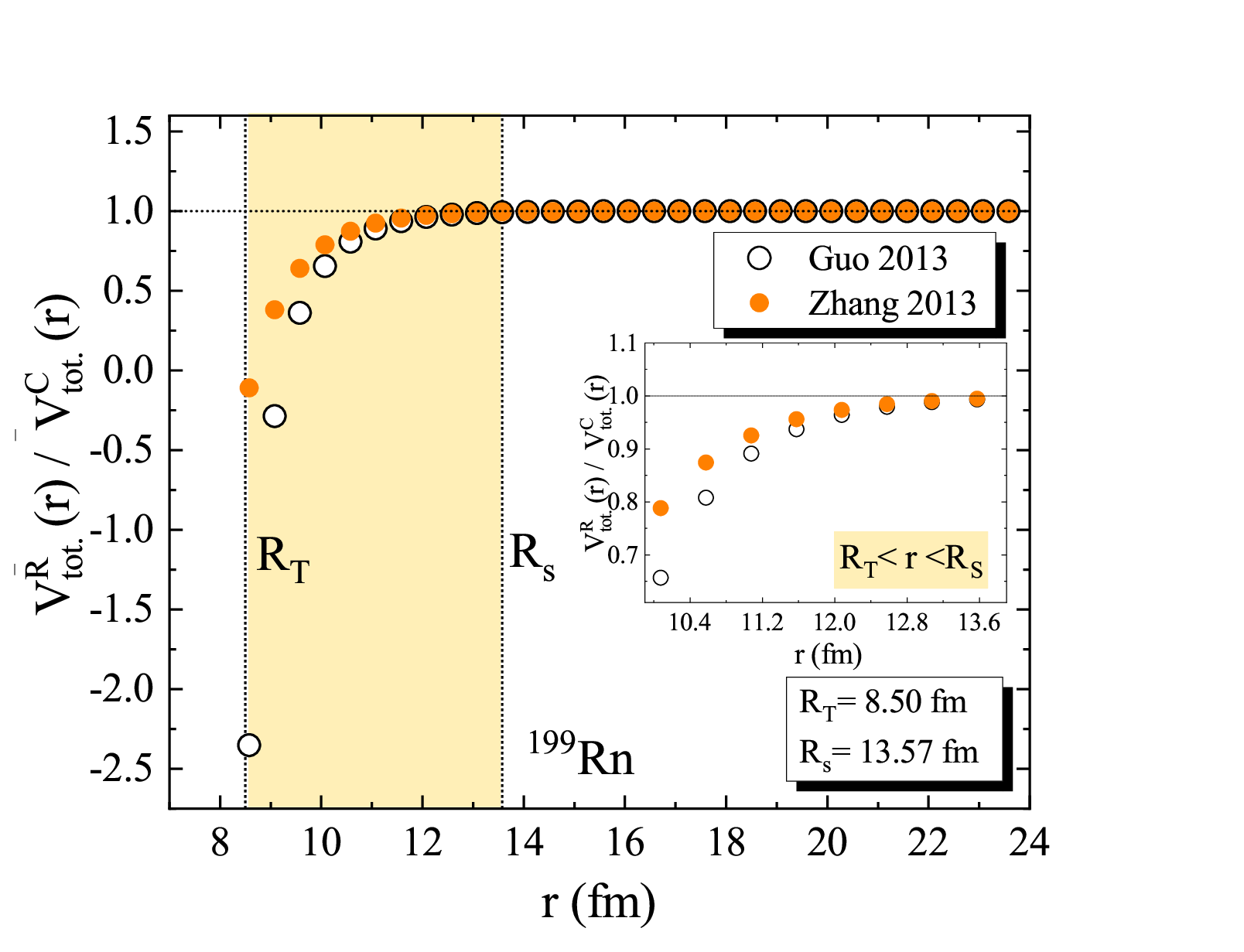}}
	\caption{\footnotesize (Colored online) The behavior of $\frac{V_{\rm tot}^{\bar{R}}}{V_{\rm tot}^{\bar{C}}}$ ratio as a function of radial distance $r$ (in fm) for $^4$He cluster ($^{199}$Rn decay). The touching point and saturation radius values are $R_T= 8.50$ fm and $R_S= 13.57$ fm, respectively, which are indicated by vertical dashed lines. The horizontal dashed line refers to the saturation value $\frac{V_{\rm tot}^{\bar{R}}}{V_{\rm tot}^{\bar{C}}}=1$.}
	\label{potenratio}
	\alttext{Description of the figure}
\end{figure}

In the proximity force theorem, the nucleus-nucleus potential is derived from the information based on the liquid drop model and general geometrical arguments. In this approach, the relationships between geometrical properties of leptodermous distributions are employed in the interpretation of the nuclear density distributions. This term is used to describe the distributions which are essentially homogeneous except at the surface (having a thin skin). In this approach, the surface width $b$ is considered as a measure of the diffuseness of the nuclear surface. Hence, the effect of nuclear surface diffuseness in the proximity form of the potential can be entered via the following definition,
\begin{equation} \label{Cbar}
	\bar{C}=\frac{C_{\alpha}C_d}{C_{\alpha}+C_d}.
\end{equation}
This equation is used instead of Eq.~(\ref{Rbar}) which is defined by the effective sharp radius $R_i$. Here, the central radius $C_i$ is introduced in the following form,
\begin{equation} \label{Cbar}
	C_i=R_i\Biggl[1-\bigg(\frac{b}{R_i}\bigg)^2\Biggl] ~~ (i=\alpha,d).
\end{equation}
S\"{u}ssmann defined the central radius $C$ as the first moment of the surface weight function $g(r)$, where $g(r)$ is the derivative of the normalized density distribution function $f(r)$ (with $f(0) = 1$), see Fig. 1 of Ref.~\refcite{29}. To investigate the role of nuclear surface diffuseness in the alpha-decay half-lives, we analyze 300 parent nuclei with $Z = 64-106$ using the Coulomb and proximity potential model. The nuclear part of the interaction potential is calculated by the Guo 2013 and Zhang 2013 proximity potentials. To calculate the centrifugal potential $V_{\ell}(r)$, it is necessary to know the value of the $\ell$ angular momentum carried away by the emitted alpha particle. The $\alpha$-particle emission from nuclei follows the spin-parity selection rule. Thus the angular momentum of the emitted $\alpha$-particle satisfies the conditions,
\begin{equation} \label{angul}
	|I_i-I_j|\leq \ell_{\alpha} \leq I_i+I_j ~~~~~~~ \frac{\pi_i}{\pi_j}=(-1)^{\ell_{\alpha}}.
\end{equation}
Here, $I_{j(i)}$, $\pi_{j(i)}$ are the spin and parity values of the parent nucleus in state $j$ (and the spin and parity values of the daughter nucleus in state $i$). Note that the $\alpha$-particle spin and parity are $\pi_{\alpha} = +1$ and $I_{\alpha} = 0$. The Q-value for the $\alpha$-decay and the experimental half-lives are evaluated using the Refs.~\refcite{30,31}. 
Here and in the following, we are interested in analyzing the behavior of $\frac{\bar{C}}{\bar{R}}$ ratio in terms of the mass number of various emitted clusters (including $^4$He, $^8$Be, $^{12}$C, and $^{14}$C) of $^{122}$294 parent nucleus. The result is shown in Fig. \ref{CbarRbar}. It is clearly observed that the $\frac{\bar{C}}{\bar{R}}$ ratio for the $\alpha$ cluster is smaller than other ones. This means that the diffuseness of the nuclear surface plays a decisive role in the alpha decay process and therefore motivates us to investigate it.
Now, we intend to investigate the behavior of the ratio of the interaction potentials in $\bar{R}$ and $\bar{C}$ forms $\frac{V_{\rm tot}^{\bar{R}}}{V_{\rm tot}^{\bar{C}}}$ as functions of the internuclear distance $r$ (fm) around the touching point (namely $R_T=R_{\alpha}+R_d$). It should be noted that the results are obtained for the radioactive decays of $^{199}$Rn by the emission of $^4$He. The results are shown in Fig.~\ref{potenratio}. From this figure, one can see that the ratio of interaction potentials has a saturation limit (when $\frac{V_{\rm tot}^{\bar{R}}}{V_{\rm tot}^{\bar{C}}}=1$) at the saturation radius $R_S$. We notice that the importance of nuclear surface diffuseness effects increases by going from the saturation radius $R_S$ to the touching point $R_T$. In fact, the obtained results reveal that the nuclear surface diffuseness effects for the alpha decay process can be important in the range of $R_T\leq$ r $\leq R_S$. As shown in Fig.~\ref{potenratio}, one can find that the prediction of the Zhang 2013 and Guo 2013 proximity potentials for the values of the saturation radius $R_S$ is almost identical. For separation distance $r \leq R_S$, the difference between these potential models becomes apparent.

\begin{figure}[th]
	\centerline{\includegraphics[width=11cm]{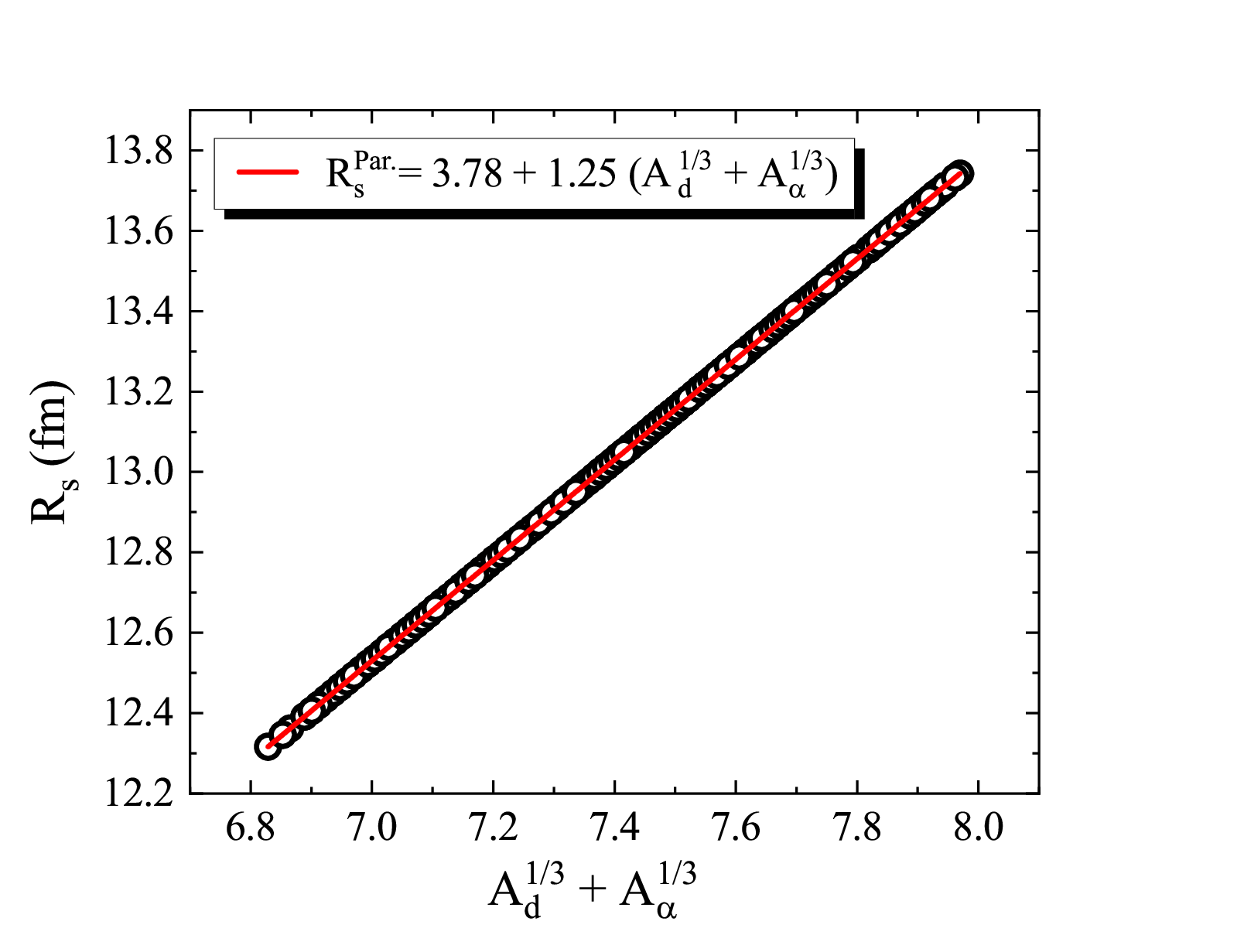}}
	\caption{\footnotesize (Colored online) The extracted values of the saturation radius $R_S$ (fm) in terms of $A_d^{1/3}+A_{\alpha}^{1/3}$ for 300 alpha decays based on the Zhang 2013 potential model. The linear fit is shown by the red line.}
	\label{RsPar}
	\alttext{Description of the figure}
\end{figure}

The behavior of saturation radius $R_S$ in terms of the $A_d^{1/3}+A_{\alpha}^{1/3}$  ($A_d$  and $A_{\alpha}$ are the mass number of daughter nuclei and alpha cluster, respectively) for 300 selected alpha decays is prepared in Fig.~\ref{RsPar}. According to this figure, one can see that the values of $R_S$ follow a regular increasing trend. Since the mass number of alpha particle is same, it is clear that the behavior of $R_S$ only depends on the daughter nuclei. We can parameterize the observed linear behavior using the following relation,
\begin{equation} \label{RsPara}
	R_s^{\rm Par.}=a+b\bigg[A_d^{1/3}+A_{\alpha}^{1/3}\bigg],
\end{equation}
where the extracted values of $(a, b)$ constants are $(3.78, 1.25)$ based on the Zhang 2013 (or Guo 2013) potential model. Fig. \ref{RsPar} shows the values of the $R_S$ for all considered decays are localized around the fitted line. Eq.~(\ref {RsPara}) enables us to estimate the radial distance at where the effects of the nuclear surface diffuseness are important in the interaction between alpha and daughter nuclei.

\begin{figure}[th]
	\centerline{\includegraphics[width=13cm]{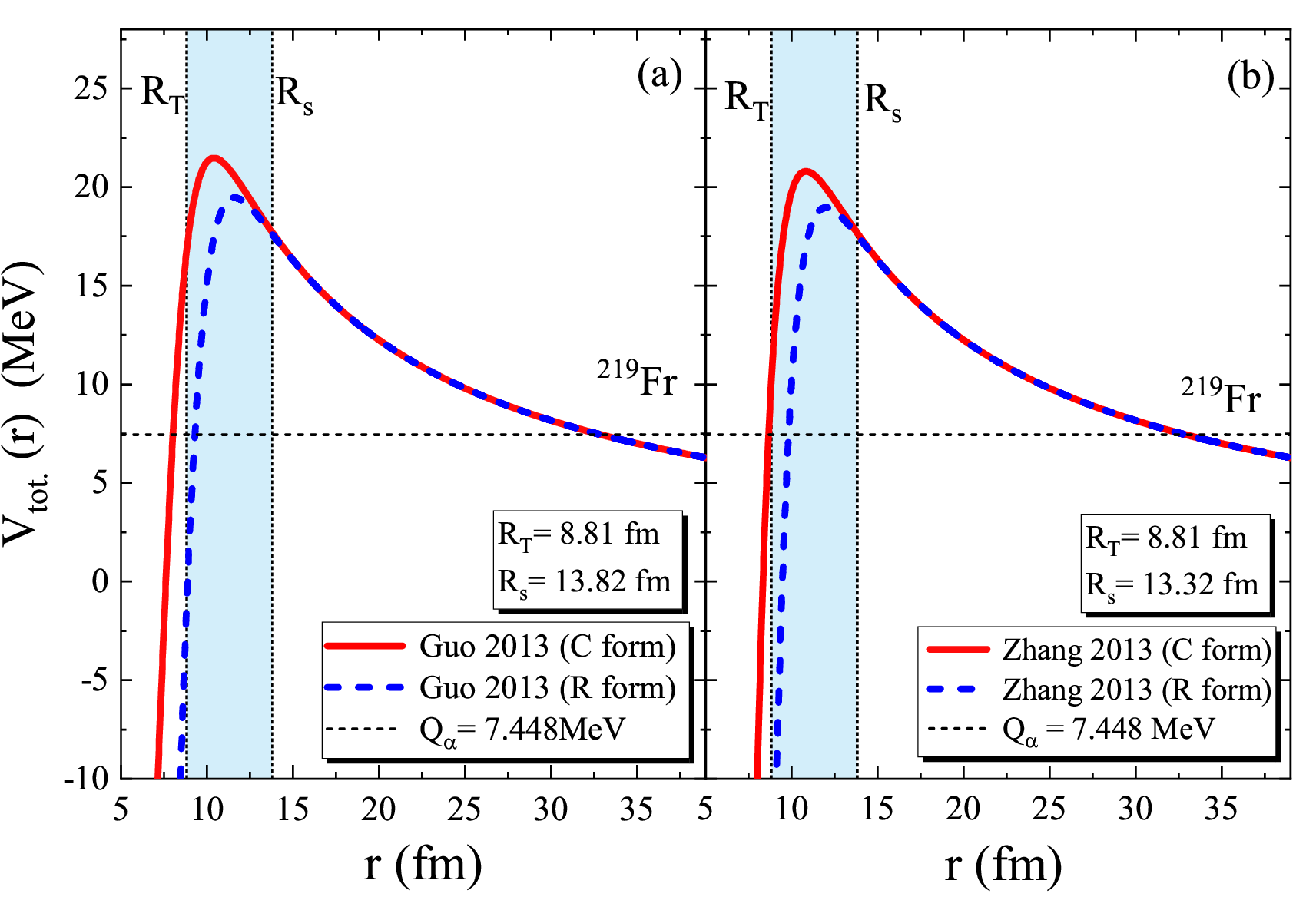}}
	\caption{\footnotesize (Colored online) The total interaction potential $V_{\rm tot} (r)$ (in MeV) as a function of the radial distance $r$ (in fm) for alpha-decay of $^{219}$Fr$\rightarrow^{215}$At+$^{4}$He using (a) Zhang 2013 and (b) Guo 2013 potential models with and without the NSD effects. The black dashed line indicates the $Q_{\alpha}$-value. The black short dotted lines show the $R_S$ and $R_T$ values.}
	\label{Poten}
	\alttext{Description of the figure}
\end{figure}

By considering the nuclear surface diffuseness effects, the nuclear potential for the selected models is modified as the following form,
\begin{equation} \label{VNmod}
	V_N (r)=4\pi \gamma b \bar{C} \Phi \bigg(\frac{r-C_{\alpha}-C_d}{b}\bigg).
\end{equation}
As can be seen from this equation, the diffuseness effects have been applied to both parts of the universal function and curvature radius. One can now calculate the nuclear potential values with and without the NSD effects. We mark the results of the Zhang 2013 potential model with and without considering these physical effects as "Zhang 2013 (C form)" and "Zhang 2013 (R form)" (Similarly, for the other potential model). In Fig.~\ref{Poten}, we plot the distributions of total emitted alpha-core interaction potential with and without the NSD effects the alpha-decay of $^{219}$Fr (with $Q_{\alpha}=7.44$ MeV), for example. From this figure, one can find that the height and thickness of the Coulomb barrier increases by considering the nuclear surface diffuseness effects. Under this situation, it is interesting to remark that the penetration probability of the alpha particle through a potential barrier decreases due to the NSD effects. In order to further understanding, the calculated values of the potential barrier height based on the different versions of the proximity potential formalisms are tabulated in Table~\ref{height}. The first and second columns denote the results of Guo 2013 proximity potential and the next two columns denote the results of Zhang 2013 model. Depending on this table, one can find that the potential barrier height increases about 2 MeV by imposing $\bar{R}\rightarrow\bar{C}$. 
In the next step, we calculate the half-life values through Eq.~(\ref{T12}).
\begin{table}[pt]
	\tbl{The barrier heights calculated using Zhang 2013 (R \& C forms) and Guo 2013 (R \& C forms) potential models for alpha decay of $^{219}$Fr, for example.\label{height}}
	{\begin{tabular}{@{}c  c c c c  c c@{}}
			\toprule  & \multicolumn{2}{c}{Guo 2013} && \multicolumn{2}{c}{Zhang 2013}\\
			\cline{1-6}
			& R form & C form && R form & C form \\
			\colrule
			$V_B$ (MeV)\hphantom{00} & \hphantom{0}18.94 & \hphantom{0}20.79 && \hphantom{0}19.45 & \hphantom{0}21.49 \\
			\botrule
	\end{tabular}}
\end{table}

\begin{figure}[th]
	\centerline{\includegraphics[width=15cm]{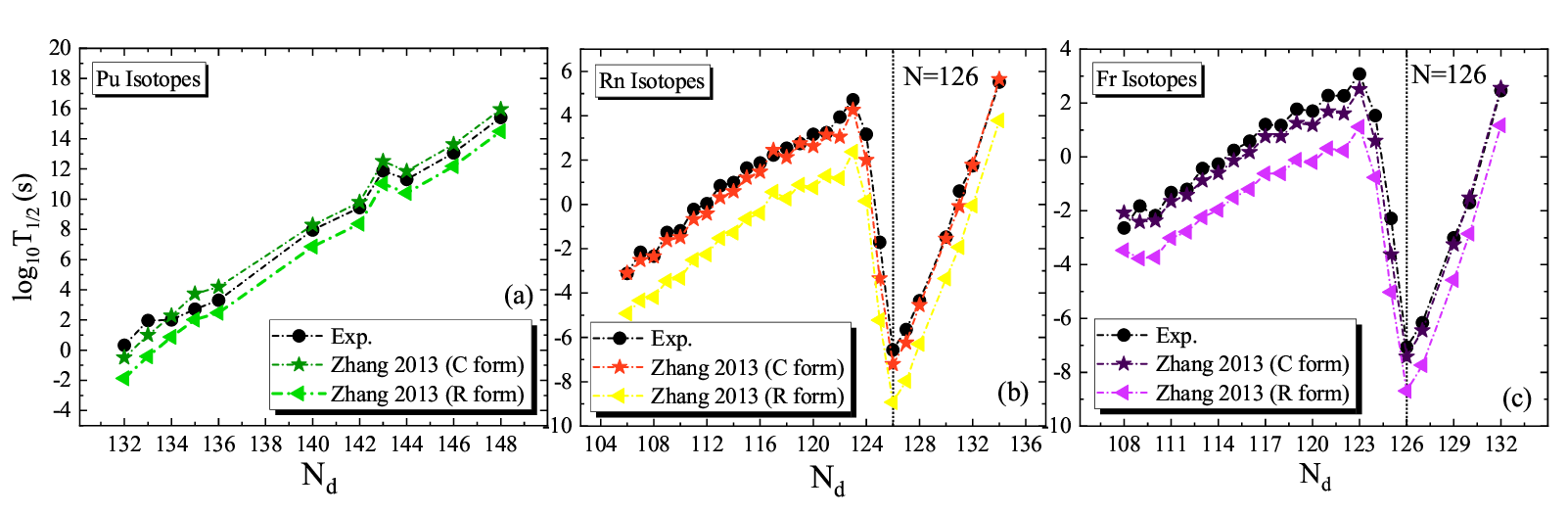}}
	\caption{\footnotesize (Colored online) Variation of the decimal logarithm of the alpha-decay half-life with the neutron number of the daughter nuclei $N_d$ for (a) Pu, (b) Rn and (c) Fr isotopic groups using Zhang 2013 (C form) (star) and Zhang 2013 (R form) (triangle) models. Experimental half-life data are shown with black circles. Also, the magic number $N=126$ is marked with a dashed line.}
	\label{logT1}
	\alttext{Description of the figure}
\end{figure}

\begin{figure}[th]
	\centerline{\includegraphics[width=15cm]{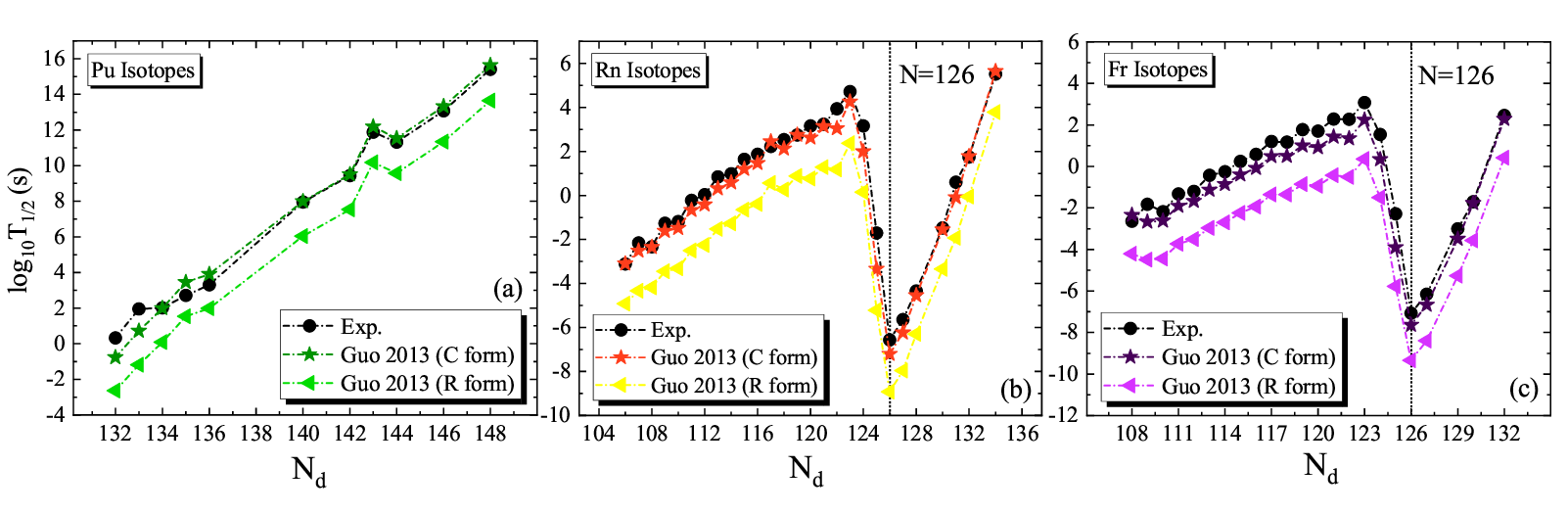}}
	\caption{\footnotesize (Colored online) Same as Fig.~\ref{logT1}, but for the Guo 2013 potential model.}
	\label{logT2}
	\alttext{Description of the figure}
\end{figure}

The logarithmic values of the half-life in terms of the neutron number of the daughter nuclei $N_d$ have been drawn in Figs.~\ref{logT1} and~\ref{logT2}, for Pu, Rn and Fr isotopic groups. The corresponding experimental data are also presented. It can be seen that the calculated logarithmic values of the half-lives in the $C$ form (star) are closer to the experimental ones (circle) than in the $R$ form.

\begin{figure}[th]
	\centerline{\includegraphics[width=9cm]{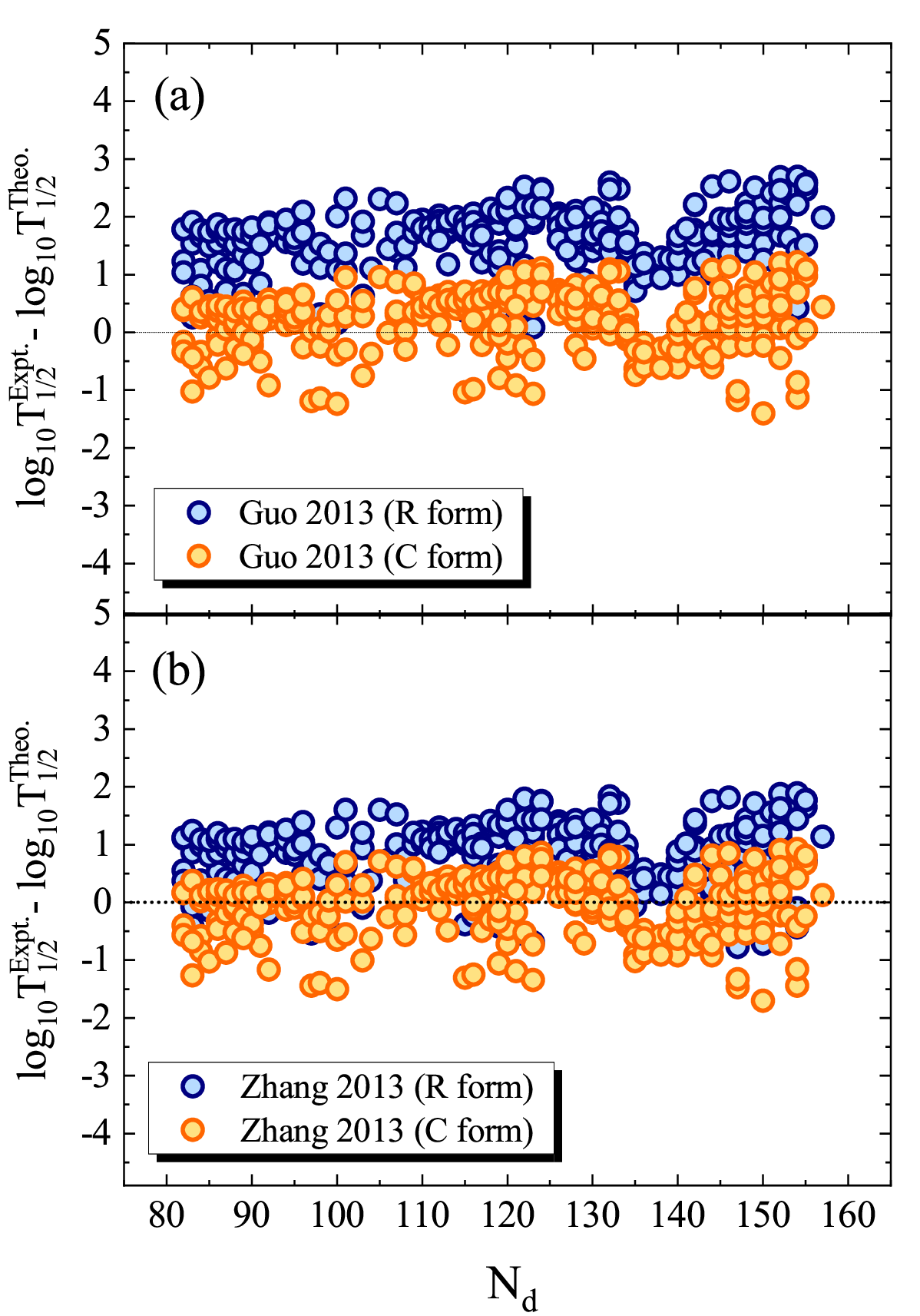}}
	\caption{\footnotesize (Colored online) The difference between the logarithm experimental and theoretical alpha decay half-lives using the original ($R$ forms: blue circles) and modified ($C$ forms: orange circles) versions of the (a) Guo 2013 and (b) Zhang 2013 models in terms of the neutron number of daughter nuclei $N_d$.}
	\label{deltaT}
	\alttext{Description of the figure}
\end{figure}

The absolute difference between the calculated values of ${\rm log}_{10}T_{1/2}$ and the corresponding experimental data within the framework of the original and modified forms of the (a) Guo 2013 and (b) Zhang 2013 models are shown in Fig.~\ref{deltaT} in terms of $N_d$. This figure shows that the NSD effects reduce the difference between the theoretical and experimental values of the $\alpha$-decay half-lives. The standard deviation $\sigma$ between the logarithm values of the theoretical alpha-decay half-lives and the corresponding experimental data can be calculated using the following expression,
\begin{equation} \label{sigma1}
	\sigma=\sqrt{\frac{1}{N}\sum_{i=1}^{N}\bigg[\rm{log}_{10}\bigg(T^{Theor.}_{1/2i}\bigg)-\rm{log}_{10}\bigg(T^{Expt.}_{1/2i}\bigg)\bigg]^2},
\end{equation}
where $N$ is the number of parent nuclei used for evaluation of the $\sigma$ value.

\begin{figure}[th]
	\centerline{\includegraphics[width=12cm]{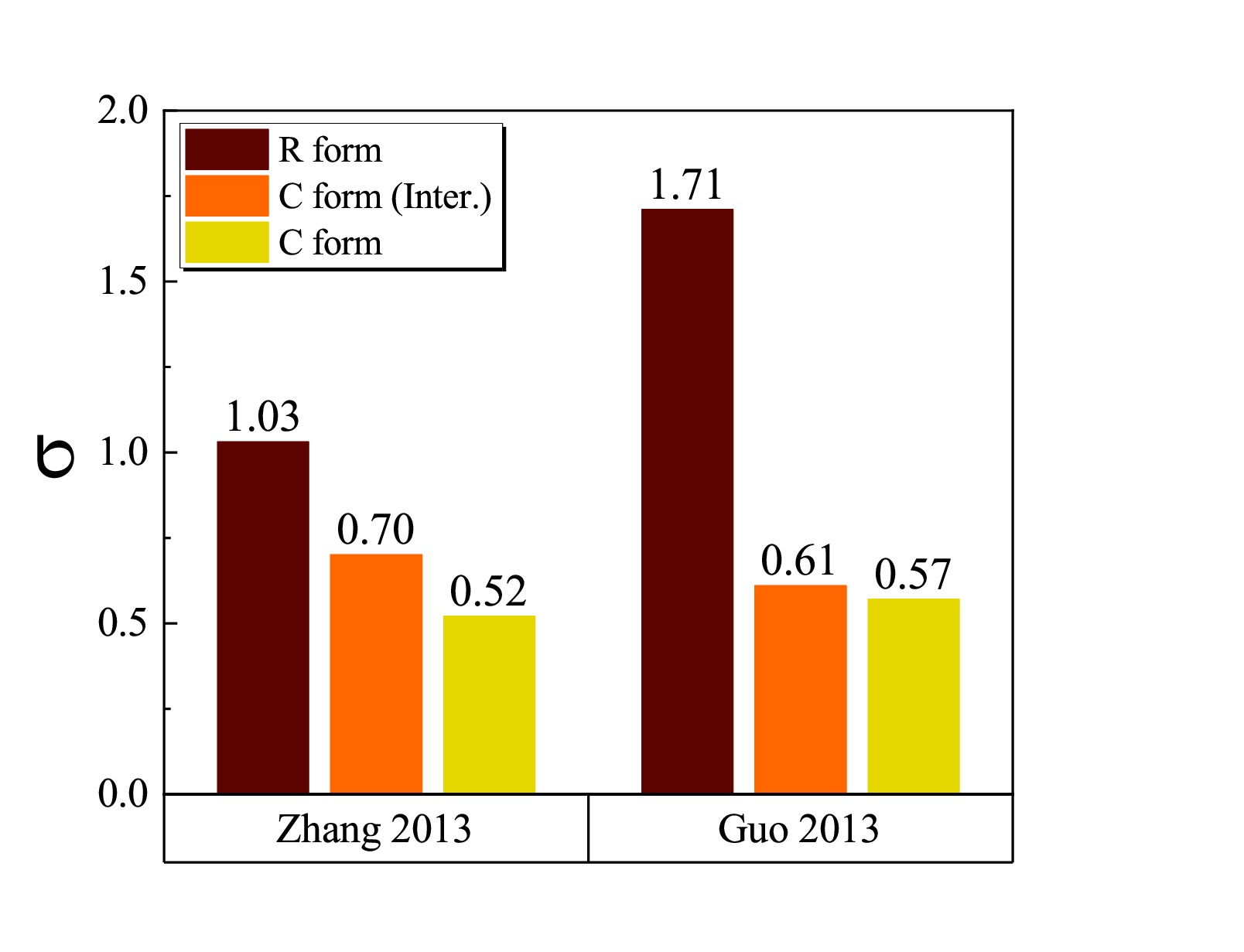}}
	\caption{\footnotesize (Colored online) Standard deviation values $\sigma$ for the studied Guo 2013 - $C$ and $R$ forms and Zhang 2013 - $C$ and $R$ form models. For comparison, the results of  the Guo 2013 and Zhang 2013 proximity potentials after imposing the diffuseness effects only through the universal function are also displayed, marked as "C form (Inter.)".}
	\label{sig}
	\alttext{Description of the figure}
\end{figure}

Here, we calculate the values of $\sigma$ using different versions of proximity potentials formalisms. The obtained values are displayed in Fig.~\ref{sig}. From this figure, one can see that the Guo 2013 (C form) and Zhang 2013 (C form) models have smaller deviations in the description of the experimental half-lives of the studied alpha emitters then the original ones. These results indicate that the agreement with experimental data enhances by imposing the nuclear surface diffuseness effects in the calculations of the total interaction potential between alpha and daughter nuclei. To gain further insight, in Fig.~\ref{sig} we also display the calculated values of the standard deviation $\sigma$ for the Guo 2013 and Zhang 2013 proximity potentials after imposing the diffuseness effects only through the universal function $\Phi(\frac{r-C_{\alpha}-C_d}{b})$. Here, we mark the results of these potentials as "C form (Intermediate)". The role of the diffuseness effects in improving the results of the original form of the proximity potentials is quite evident. However, the best agreement with the experimental data can be obtained after simultaneously applying the diffuseness effects through the universal function and curvature radius.
\subsection{Alpha decay half-lives in the SHN region}
In previous section, we found that the Zhang 2013 (C form) and Guo 2013 (C form) are able to reproduce the experimental data of alpha decay half-lives in the range of heavy nuclei. Hence, we decided to calculate the alpha decay half-lives of 50 parent nuclei in the SHN region with $Z=100-118$ and compare the results with corresponding experimental data and other theoretical approaches such as MGLDM\cite{32}, Royer\cite{33}, AKRE\cite{34}, NewRenB\cite{35}, and Akrawy\cite{36}. The results of the standard deviation $\sigma$ are reported in Table~\ref{sigm}. We previously indicated that the Zhang 2013 (C form) is a more suitable model for calculating the half-lives of alpha decay in the $Z \leq 106$ region. While, it is clearly evident from this table that the modified form of the Guo 2013 proximity model performs well in the SHN region. In addition, we find that the results produced by the Guo 2013 (C form) are comparable with those obtained from the other approaches mentioned above.
\begin{table}[pt]
	\tbl{Comparison of the calculated standard deviation $\sigma$ of the Zhang 2013 (C form), Guo 2013 (C form) and MGLDM models and Royer, AKRE, New RenB, and Akrawy formulas.\label{sigm}}
	{\begin{tabular}{@{}cccccccccccccccc@{}}
			\toprule
			Zhang 2013  & Guo 2013 & MGLDM & Royer & AKRE & NewRenB & Akrawy \\
			(C form)  &(C form)&   & & \\
			\colrule
			1.07\hphantom{00} & \hphantom{0}0.63 & \hphantom{0} 0.94 & \hphantom{0} 0.57 & \hphantom{0} 0.54 & \hphantom{0} 0.37 & \hphantom{0} 0.34 \\
			\botrule
	\end{tabular}}
\end{table}

\begin{figure}[th]
	\centerline{\includegraphics[width=13.5cm]{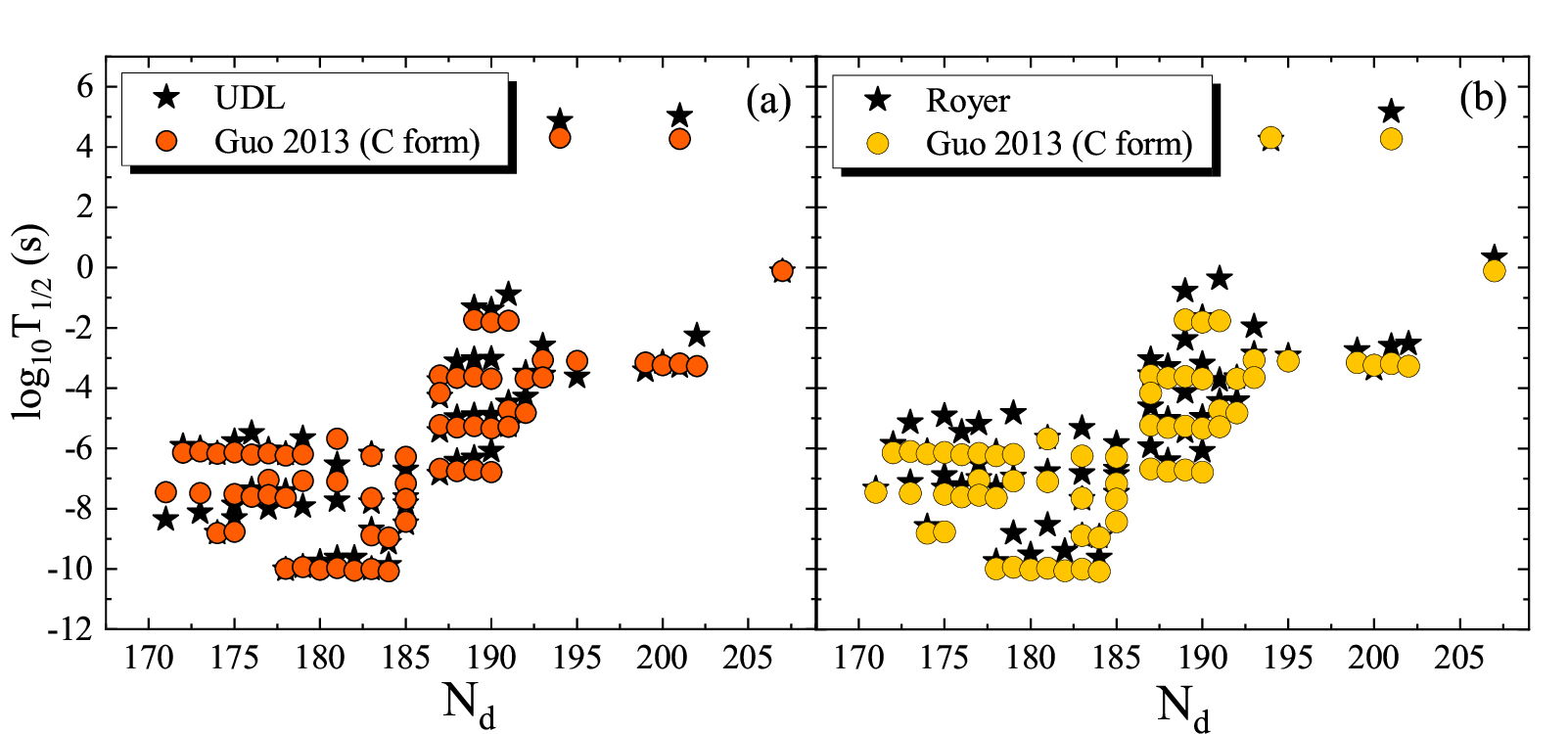}}
	\caption{\footnotesize (Colored online) The logarithmic behavior of the calculated alpha decay half-lives ${\rm log}_{10}T_{1/2}$ in terms of neutron number of daughter nuclei $N_d$. The obtained data using Royer and UDL models are represented by a star symbol and the calculated data using Guo 2013 (C form) are shown by a circle symbol.}
	\label{logTSHN}
	\alttext{Description of the figure}
\end{figure}

In the next step, we try to predict the alpha decay half-lives of the super-heavy parent nuclei. We calculated the half-lives of 65 selected nuclei in the range of super-heavy nuclei $Z=120-126$ using the Guo 2013 (C form) model. We used the Royer\cite{33} and UDL\cite{37,38} models for a comparison. The results are listed in Table~\ref{TlogT}. The first, second and third columns refer to the mass number, atomic number and released energy $Q_{\alpha}$-values, respectively. The fourth, fifth and sixth columns represent the calculated half-lives using UDL, Royer and Guo 2013 (C form), respectively. It can be seen that the calculated half-lives through the present studied model is in the order of the calculated half-lives through the UDL and the Royer models. The logarithmic behavior of the calculated alpha decay half-lives in terms of neutron number of daughter nuclei $N_d$ is displayed in Fig.~\ref{logTSHN}. The results show that the calculated half-lives (circle symbol) are within the range of Royer and UDL data (star symbol). In this situation, one can expect that Guo 2013 (C form) to be suitable in further investigating in the super-heavy mass region.

\begin{table}[pt]
	\tbl{Comparison of theoretical predictions of $\alpha$-decay half-lives through Guo 2013 (C form) model with calculated half-lives using the Royer~\cite{33} and UDL~\cite{37,38} formulas. \label{TlogT}}
	{\begin{tabular}{ccccccccccccccc}
			\toprule
			&&& \multicolumn {3}{c} {${\rm log}_{10}T_{1/2}(s)$}
			&&&&&& \multicolumn {3}{c} {${\rm log}_{10}T_{1/2}(s)$}\\
			\cline{4-6}
			\cline{12-14}\\
			\colrule
			A & Z & $Q_{\alpha}$ & UDL  & Royer & Guo 2013    &&& A & Z & $Q_{\alpha}$ & UDL  & Royer & Guo 2013 \\
			&& (MeV) &&& (C form) &&&&&(MeV) &&&(C form)  \\
			\colrule 
			307 & 120 & 13.52 & -6.71 & -5.83 & -6.28    &&& 315 & 124 & 13.21 & -4.91 & -4.12 & -5.25 \\      
			294 & 120 & 13.24 & -5.92 & -5.84 & -6.14    &&&  319 & 124 & 12.18 & -2.60 & -1.96 & -3.65 \\      
			323 & 120 & 9.12 & 5.03 & 5.17 & 4.27        &&& 313 & 124 & 13.47 & -5.43 & -4.61 & -5.22 \\      
			295 & 120 & 13.27 & -6.00 & -5.14 & -6.10    &&& 317 & 124 & 13.00 & -4.48 & -3.73 & -5.28 \\      
			296 & 120 & 13.34 & -6.17 & -6.08 & -6.17    &&& 300 & 124 & 15.34 & -8.81 & -8.58 & -8.81 \\      
			316 & 120 & 9.19 & 4.87 & 4.23 & 4.31        &&& 305 & 124 & 14.80 & -7.92 & -6.93 & -7.07 \\      
			313 & 120 & 11.02 & -0.88 &  -0.37 & -1.76   &&& 314 & 124 & 13.24 & -4.97 & -5.02 & -5.30 \\      
			312 & 120 & 11.22 & -1.40 & -1.64 & -1.81    &&& 309 & 124 & 15.19 & -8.70 & -7.68 & -8.88 \\      
			297 & 120 & 13.14 & -5.76 & -4.92 & -6.13    &&& 311 & 124 & 14.70 & -7.83 & -6.86 & -7.16 \\      
			298 & 120 & 13.01 & -5.49 &  -5.45 & -6.21   &&& 318 & 124 & 12.56 & -3.49 & -3.65 & -3.69 \\      
			305 & 120 & 13.28 & -6.18 & -5.32 & -6.25    &&& 310 & 124 & 15.43 & -9.14 & -8.93 & -8.96 \\      
			311 & 120 & 11.20 & -1.33 &  -7.75 & -1.73   &&& 313 & 126 & 15.37 & -8.49 & -7.47 & -8.43 \\      
			299 & 120 & 13.26 & -6.04 &  -5.18 & -6.16   &&& 316 & 126 & 14.23 & -6.42 & -6.40 & -6.76 \\      
			301 & 120 & 13.06 & -5.66 & -4.83 & -6.19    &&& 308 & 126 & 16.16 & -9.77 & -9.52 & -11.00 \\     
			300 & 120 & 13.32 & -6.18 & -6.11 & -6.24    &&& 317 & 126 & 14.17 & -6.32 & -5.44 & -6.71 \\      
			309 & 120 & 12.16 & -3.74 & -3.04 & -4.16    &&& 307 & 126 & 16.27 & -9.92 & -8.81 & -9.94 \\      
			312 & 122 & 12.16 & -3.11 & -3.27 & -3.65    &&& 311 & 126 & 16.28 & -10.00 & -8.89 & -10.00 \\    
			313 & 122 & 12.13 & -3.04 & -2.38 & -3.61    &&& 310 & 126 & 16.06 & -9.63 & -9.40 & -10.00 \\     
			302 & 122 & 14.24 & -7.42 & -7.29 & -7.63    &&& 309 & 126 & 16.08 & -9.64 & -8.54 & -9.97 \\      
			309 & 122 & 14.28 & -7.62 & -6.68 & -7.67    &&& 319 & 126 & 13.64 & -5.24 & -4.44 & -4.73 \\      
			295 & 122 & 14.80 & -8.35 & -7.33 & -7.46    &&& 318 & 126 & 14.05 & -6.09 & -6.09 & -6.79 \\      
			300 & 122 & 14.22 & -7.36 & -7.22 & -7.60    &&& 320 & 126 & 13.19 & -4.29 & -4.40 & -4.81 \\      
			305 & 122 & 13.76 & -6.54 & -5.65 & -5.68    &&& 321 & 126 & 12.86 & -3.56 & -2.86 & -3.07 \\      
			311 & 122 & 12.67 & -4.29 & -3.55 & -3.58    &&& 306 & 126 & 16.34 & -10.00 & -9.74 & -9.99 \\     
			301 & 122 & 14.26 & -7.45 & -6.50 & -7.55    &&& 315 & 126 & 14.46 & -6.85 & -5.94 & -6.68 \\      
			307 & 122 & 14.39 & -7.79 & -6.83 & -7.64    &&& 323 & 126 & 12.87 & -3.63 & -2.92 & -3.10 \\      
			297 & 122 & 14.65 & -8.12 & -7.12 & -7.49    &&& 327 & 126 & 12.76 & -3.42 & -2.74 & -3.16 \\      
			299 & 122 & 14.50 & -7.88 & -6.89 & -7.52    &&& 328 & 126 & 12.64 & -3.15 & -3.36 & -3.23 \\      
			314 & 122 & 12.12 & -3.03 & -3.20 & -3.68    &&& 312 & 126 & 16.19 & -9.87 & -9.62 & -10.01 \\     
			316 & 124 & 13.20 & -4.89 & -4.95 & -5.33    &&& 329 & 126 & 12.68 & -3.28 & -2.61 & -3.19 \\      
			307 & 124 & 14.68 & -7.74 & -6.76 & -7.10    &&& 330 & 126 & 12.26 & -2.26 & -2.53 & -3.26 \\      
			301 & 124 & 15.05 & -8.32 & -7.30 & -8.76    &&& 335 & 126 & 11.41 & -0.12 & 0.33 & -0.10 \\       
			303 & 124 & 14.58 & -7.99 & -6.99 & -7.04\\                            
			\botrule
	\end{tabular}}
\end{table}
\section{Conclusions}\label{4}
We used 300 parent nuclei with $Z=64-106$ to study the effects of nuclear surface diffuseness on the alpha decay characteristics. For this purpose, we employ Zhang 2013 and Guo 2013 models to calculate the nuclear interaction potential. The nuclear surface diffuseness effects are applied in the calculations using the reduced radius $\bar{C}$. By analyzing the radial behavior of the total interaction potential for different $\alpha$-nuclei systems, one can find that the mentioned physical effects increase the potential barrier height. In this paper, we present an investigation to explore the effect of shell closure (around the magic number $N=126$) on the alpha decay half-lives through the plot of logarithmic values of $T_{1/2}$ evaluated using different nuclear potentials versus neutron number of the daughter nuclei. When compared to available experimental data, there are improvements in the results when the interaction potentials with NSD effects are used compared to when the NSD effects are not employed. For further understanding, in this work, we calculate the standard deviations between the logarithmic values of alpha radioactivity half-lives of calculations and experimental data within two theoretical models. Our results reveal that the calculated values of $\sigma$ based on the Zhang 2013 and Guo 2013 potential models with the NSD effects ($\sigma^{\rm Zhang ~2013~(C ~form)}= 0.52$ and $\sigma^{\rm Guo~2013~(C~form)}= 0.57$) are smaller than those obtained using their original versions ($\sigma^{\rm Zhang~2013~(R~form)}= 1.03$ and $\sigma^{\rm Guo~2013~ (R ~form)}= 1.71$). The difference between the $\sigma$ values of Zhang 2013 (C form) and Guo 2013 (C form) models can be attributed to the different in their universal functions. In addition, we calculated the half-lives for 50 super-heavy nuclei (SHN) using Zhang 2013 (C form) and Guo 2013 (C form) models. The obtained half-lives compared with experimental data and also with existing theoretical models such as Royer, AKRE, New RenB, and Akrawy formulas. The comparison indicated that the Guo 2013 (C form) model is a useful model to investigate the alpha decay half-lives of SHN. Then the predictions of alpha decay half-lives for 65 SHN with $Z = 120-126$ were made by using Guo 2013 (C form) model. It was found that there is a good agreement between our predicted half-lives and those obtained from the semi-empirical formulas such as Royer and UDL.

\end{document}